# SAP HANA Data Volume Management

Subhadip, Kumar*

Western Governors University, skuma19@wgu.edu

Today's information technology is a data-driven environment. The role of data is to empower business leaders to make decisions based on facts, trends, and statistical numbers. SAP is no exception. In modern days many companies use business suites like SAP on HANA (S/4 or ERP) or SAP Business Warehouse and other non-SAP applications and run those on HANA databases for faster processing. While HANA is an extremely powerful in-memory database, growing business data has an impact on the overall performance and budget of the organization. This paper presents best practices to reduce the overall data footprint of HANA databases for three use cases – SAP Business Suite on HANA, SAP Business Warehouse, and Native HANA database.

**Additional Keywords and Phrases:** SAP, HANA, NSE

## 1 INTRODUCTION

Many organizations adopt HANA (High-performance ANalytic Appliance) as primary database for SAP and non-SAP application over traditional databases (Oracle/SQL Server) because of its in-memory computing and real time results[3]. It retrieves data 3600 times faster than traditional databases and can scan up to 3.5 billion records per second per core.

By design HANA is a multi-model database that stores data in its memory instead of keeping it on disk. HANA utilizes column store mechanism to store the data on the tables. Memory on SAP HANA database is divided in two parts – actual table data and working memory. SAP typically recommends to maintain 1:1 ratio between table data and workspace memory. Increasing table data will also increase workspace memory requirement. SAP HANA runs on certified hardware either on-premise or cloud. Typically, memory configuration runs from 256GB and all the way up to 12TB in a scale up architecture with matching CPU configuration[4].

It is very important to properly size the database to accommodate current data and future growth. Upgrading memory of existing HANA databases are not only complicated but also very expensive. Most of the mission critical application that run on HANA usually configured in a HA - high availability (same data center) and DR - disaster recovery (different data center) setup. In a typical example where a mission critical HANA database grew from 1TB to 2TB, memory and CPU has to be upgraded in 3 places – 2 HA and 1 DR.

In some instances, entire physical server has to be replaced in order to accommodate the growth. For example a database which is currently running on a 6TB/4 socket hardware on Cisco C480 and now requires 8TB memory due to data growth – it is not possible just to upgrade the memory and CPU module but a new upgraded hardware C890 with 12TB/8 socket has to be procured. Setting up new hardware not only take up space in a datacenter but also storage, power supply, networking, OS installation, maintenance and setup need to complete before HANA can be installed and existing database can be migrated to new hardware. All these activities having a direct impact on Total Cost of Ownership (TCO). This also have a direct impact on the overall green day initiative of the organization[8].

---

* Place the footnote text for the author (if applicable) here.

Therefore, it is extremely important to control the database growth to control the TCO. Fig 1 shows a pie chart of TCO distribution.

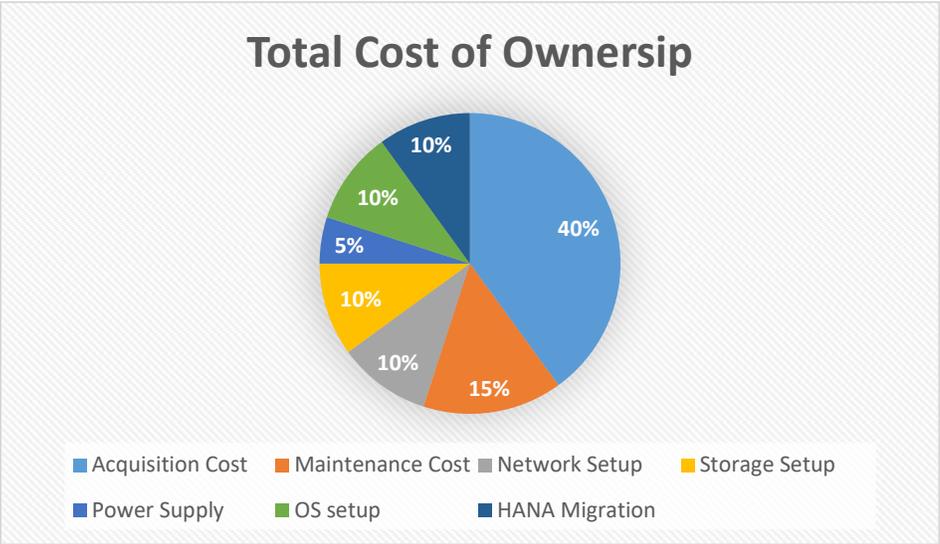

Fig 1. Total Cost of Ownership

## 2 DATA TIERING - BASICS

In this section overview of multi temperature data tiering and methods to implement that will be explained

### 2.1 Multi Temperature Data and Data Tiering

HANA database data can be managed according to how frequently it is accessed. This sometimes referred as Multi-temperature database management. It gives you the ability to keep mission critical data to HOT layer i.e. in memory and move infrequently access data to WARM layer i.e. to disk and rarely used data to an inexpensive storage solution like Hadoop.

For an example, finance team requires to generate a daily revenue report for C-level executives and therefore they access last 48hrs of data very frequently which can reside in HOT layer (0 – 48hrs). They also need to generate a monthly, quarterly and yearly report and need to pull last 30, 90 and 365 days of financial data once in a while which can reside in WARM layer (48hrs to 365 days). Also finance team needs to retain 7 years of data for audit purpose which but it has been rarely accessed can be stored in COLD layer (1year to 7 year). Fig 2 below represents significance of each layer.

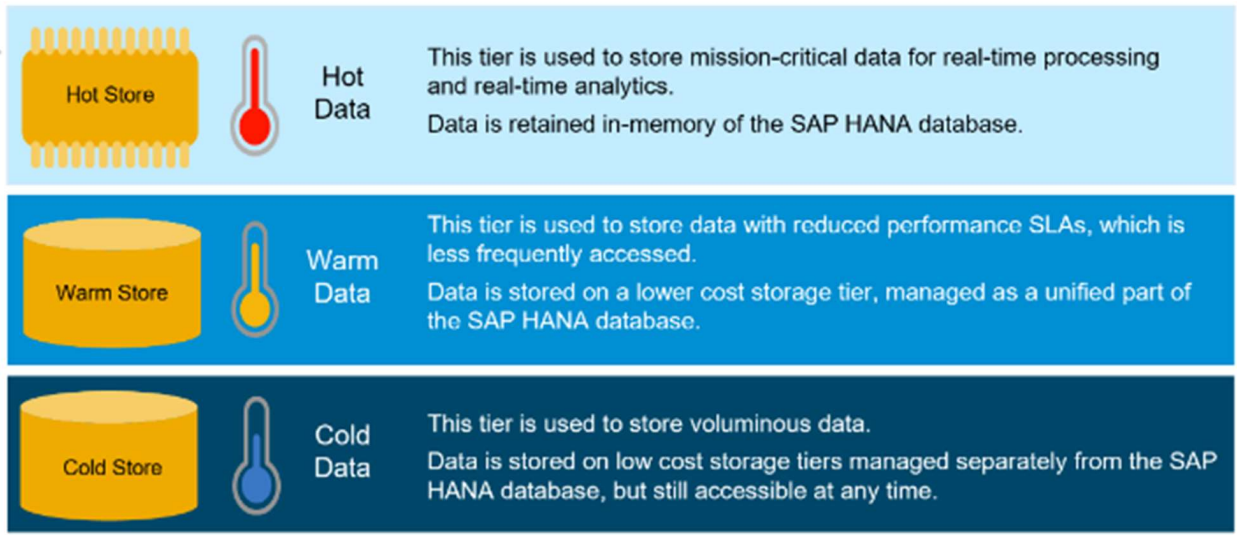

Fig 2. Multi Temperature Data-Tiering

Hot layer:

- Very fast data retrieval
- Frequently accessed data
- High Cost due to in-memory in nature
- Technology: DRAM, PMEM

Warm Layer:

- Very fast data retrieval
- Infrequently accessed data
- Low Cost compared to Hot Layer as it resides in disk
- Technology: HANA NSE (Native Storage Extension), Dynamic Tiering, Extension Node, Data Aging

Cold Layer:

- Slow retrieval of data
- Rarely accessed data
- Lowest cost

Technology – NLS (Near Line Storage), SAP IQ, SAP ILM, SAP DWF (Data Warehousing Foundation).

## 2.2 Data Tiering Methodology

It is confusing which option to choose for data tiering in HANA[10]. Unfortunately, not one-size-fits-all in HANA – it finally boils down to whether and what business suites on top of HANA or whether it is a standalone native HANA database. Figure 3 explains different methodologies that can be used for each layer of data.

|       | **Native SAP HANA** | **SAP BW on HANA or SAP BW/4 HANA** | **SAP Business Suite on HANA or SAP S/4 HANA** |
|-------|---------------------|--------------------------------------|------------------------------------------------|
| **HOT** | PMEM or DRAM | PMEM or DRAM | PMEM or DRAM |
| **WARM** | SAP HANA dynamic tiering | SAP HANA extension node | Data aging |
|       | SAP HANA extension node | Native Storage Extension (BW/4 HANA only) | SAP HANA NSE |
|       | Native Storage Extension |  |  |
| **COLD** | Data Lifecycle Manager (DLM) with SAP Data Hub | SAP BW NLS with SAP IQ | ILM Store with SAP IQ |
|       | DLM with SAP HANA Spark Controller | SAP BW NLS with Hadoop and Amazon S3 | Data archiving |
|       |  | SAP BW/4 HANA Data Tiering Optimization (DTO) with SAP Data Hub and Cloud |  |

Fig 3. Data Tiering Technologies

## 3 DATA TIERING – NATIVE SAP HANA

### 3.1 Hot Data

For all HANA database whether it is a native HANA DB or SAP Business Suite on HANA – hot layer will always be the DRAM (Dynamic Memory) or PMEM (Persistent Memory) [7]. HANA stores columnar data in memory for faster processing. Use of Optane aka PMEM is decreasing over time both on premise and cloud. Except Microsoft Azure no other cloud providers offer PMEM based hardware. Intel recently announced to winding down its Optane business.

## 3.2 Warm Data

*3.2.1 Warm Data*

HANA extension node is introduced on HANA 2.0 SP03 for native HANA database. Extension node is based on the HANA scale-out feature. In this architecture one worker node (slave node) in the scale-out landscape is reserved for warm-data storage and processing. Extension node allows larger data footprint by default 100% of the node DRAM size (optional: 200%). Extension node is also having a relaxed core/memory ratio with HANA TDI5.

In order to accommodate larger data footprint which is up to 200% of DRAM, reconfiguration on storage I/O level, table partitioning on extension node required in order to accommodate 200% of warm data. It degrades the performance[6] of extension node because of increased disk access and unload/reload.

Advantages:

- Easy to implement and manage as it is based on HANA scale-out mechanism
- It gives almost same in-memory performance
- Full functional parity with HANA database
- It stores WARM data upto 100% of DRAM and optionally 200%
- Multiple extension node possible but not comes by default

Disadvantages:

- Higher TCO
- It can only store WARM data up to 100% of DRAM and optionally 200%
- It requires HANA TDI5 certified hardware

*3.2.2 SAP HANA Dynamic Tiering*

HANA Dynamic Tiering is another option for WARM data storage for managing less frequently used data. This is based on disk-centric technology where columnar data resides on disk. SAP HANA dynamic tiering exists within the SAP HANA system architecture as a dedicated database process, named esserver. Like the indexserver process, which stores and processes in-memory data, the esserver process stores data in columnar, disk-based structures and offers disk-optimized data processing.

In non-production HANA environment, esserver can be co-deployed on the same host as SAP HANA for scale up architecture. For production environments, SAP recommends dedicated host for essserver. For scale-out systems, ess server should still be installed on its own machine.

With multiple tenant databases, a dedicated esserver process and dynamic tiering extended store is required for each tenant database using dynamic tiering. Currently, dynamic tiering does not support high tenant isolation.

Even though it is possible to add HANA dynamic tiering for small databases however SAP recommends dynamic tiering for databases larger than 512GB or larger where large data volumes begin to necessitate a data lifecycle management solution.

The recommended ratios of SAP HANA memory to SAP HANA dynamic tiering extended storage are:

- SAP HANA memory <= 2.5TB: size of dynamic tiering storage should not exceed 4x the size of SAP HANA memory.

- SAP HANA memory > 2.5TB: size of dynamic tiering storage should not exceed 8x the size of SAP HANA memory.

HANA Dynamic Tiering host memory requirement is much smaller than HANA in-memory host. Fig 5 represents a table for memory requirement of HANA Dynamic Tiering host by GCP (Google Cloud Platform).

| Dynamic tiering data capacity | Memory |
|---|---|
| 512 GB | 52 GB |
| 1,300 GB | 104 GB |
| 5,000 GB | 208 GB |
| 10,000 GB | 1433 GB |

Fig 4: Dynamic Tiering Memory GCP

Advantages:

- This applies to both SAP scale out and scale up architecture
- Low TCO
- No requirement of TDI certified hardware
- No separate license required to implement HANA Dynamic Tiering

Disadvantages:

- In certain circumstances, HANA dynamic tiering doesn't support an operation where entire dataset being transferred from dynamic tiering to SAP HANA and HANA host doesn't have sufficient memory to perform that.
- HANA Dynamic Tiering is slow compared to Extension Node

3.2.3 Native Storage Extension

SAP HANA Native Storage Extension (NSE) [2] is a general-purpose, built-in warm data store in SAP HANA that lets you manage less-frequently accessed data without fully loading it into memory. It integrates disk-based or flash-drive based database technology with the SAP HANA in-memory database for an improved price-performance ratio. This solution is available from HANA 2.0 SP04 onwards.

By default all HANA columnar tables are Column Loadable that means they load on the memory for better performance – using NSE you can convert a table to page loadable which means data from table will be loaded in memory in granular units of pages for query processing, remaining pages will reside in disk.

For NSE you can add warm storage upto 1:4 ratio of HANA hot data in memory to warm data on disk. NSE disk should be no larger than 10TB. Dynamic Tiering is having much larger capacity.

SAP HANA Native Storage Extension (NSE) Advisor in SAP HANA to get suggestions on load units for tables, partitions, or columns based on how frequently they are accessed. It determines the temperature of data and uses rule-based heuristics to identify hot and warm objects as candidates to be either page-loadable or column-loadable. A rule-based (data temperature threshold, access pattern and frequency, and data density) algorithm is used to derive these recommendations.

Main component of NSE is buffer cache. Which is required for performance access to pages on disk - With the current NSE feature by default, 10% of the main memory is reserved for buffer cache and Not allocated. The Buffer Cache uses LRU (Last Recently Used) and HBL (Hot Buffer List) strategies and reuses the pages from the internal pools instead of allocating/deallocating pages via HANA memory management.

Advantages:

- No separate component needs to be installed
- Low TCO
- More granularity – NSE can be enabled table level, column level, partition level
- No separate license required to implement HANA NSE
- NSE can be used both scale-up and scale-out architecture
- No separate hardware is needed

Disadvantages:

- NSE advisor requires setup of buffer cache, which inturn requires NSE advisor and few round of iteration to determine the candidates for NSE
- HANA Dynamic Tiering is slow compared to Extension Node

In contrast to Dynamic Tiering, the query execution in the HANA service, storing the NSE data, creates transient data and interim result in memory only. Thus, the memory requirement for a comparable workload can be higher with NSE. A solution to migrate data from Dynamic Tiering to NSE in on the road map for SAP HANA.

### 3.2.4 SAP Data Aging

Data aging offers the option of moving large amounts of data within a database to get more working memory. The SAP application helps you move data from the current area to the historical area. The application controls the move by specifying a data temperature for the data. You can influence the move by an aging-object-specific customizing, usually using a residence time. Moving the data influences the visibility of the data during data access [1].

Data aging exercise has to be performed at ABAP layer.

High Level steps are:

- Create and manage partitions
- Activate the Data Aging object
- Define residence time for Data Aging
- Create and manage data aging group
- Schedule Data Aging runs

Advantages:

- No separate license is required
- Low TCO
- No separate hardware is needed

Disadvantages:

- Only applicable to certain ABAP business suite

- Only applicable to predefined data aging objects – any new object that is not part of standard data aging object needs a custom development which is time consuming

### 3.3 Cold Data

Cold data refers to the data that is seldom or sporadically accessed. Separating cold data from the SAP HANA database reduces the database footprint with tables or partitions moved from SAP HANA to external storage with mostly read-only data access and separate high availability, disaster recovery, encryption, and admin functionality.

*3.3.1 DLM with DataHub/Spark Controller*

There are two approaches to access SAP HANA cold storage: SAP Data Hub and SAP Spark Controller.

SAP Data Hub deployed in Kubernetes cluster, SAP Data hub distributed runtime engine (also known as Vora) can persist cold data in disk-based, streaming tables. Technically, these streaming tables are viewed as Virtual tables by SAP HANA. SAPHANA queries those virtual tables from the SAP Data Hub using Vora ODBC Adapter with SAP HANA Smart Data Access (SDA). Data Lifecycle Management tool (DLM) of SAP Data Warehouse Foundation (DWF) facilitates the bi-directional movement of data between hot, warm and cold layer. For external storage in this option we can either utilize on premise HDFS or cloud based S3, Azure Data Lake Storage (ADLS) or GCP.

Another option is to use HANA Spark Controller which allows SAP HANA to access Hadoop [9] data through SQL interface and primarily works with Spark SQL to connect to an existing Hive metastore. It uses SparkSQL smart data access (SDA) Adapter which moderates query execution and data transfer by enabling SAP HANA to fetch data in a compressed columnar format.

Advantages:

- Faster access of data
- Portability – Any underlying external storage can be used (on-premise and cloud)

Disadvantages:

- Requires separate license
- Main use case is Native HANA DB
- Complex architecture that requires early planning

*3.3.2 Data Archiving*

In an operative SAP BW system, the volume of data increases constantly because of business and legal requirements. The large volume of data can affect the performance of the system and increase the administration effort, which results in the need to implement a data-aging strategy.

If you want to keep the amount of data in your SAP BW system without deleting, you can use data archiving. The data is first moved to archive or near-line storage and then deleted from the SAP BW system. You can either access the data directly or load it back as required, depending on how the data has been archived.

Near-line storage (NLS) is used to archive Business warehouse (BW) data which can still be available for reporting using Bex queries. Once data is archived to NLS, BW still needs some adjustments in order to effectively report data from NLS. Once data is archived to NLS, BW still needs some adjustments in order to effectively report data from NLS. It mainly uses SAP IQ as database for the NLS solution.

Advantages:

- SAP NLS based on SAP IQ works with all traditional databases as well as SAP HANA
- SAP IQ being a SAP product support for SAP NLS is robust

Disadvantages:

- Requires separate license
- Main use case is SAP BW or BW/4HANA
- Slower retrieval

*3.3.3 Data Archiving*

This is traditional data archiving process at ABAP layer using tcodes SARA, SARI, TAANA[5]. It is only applicable for standard archiving objects for OLTP systems like SAP ECC and SAP S/4 HANA. In this process data that is being archived stores as a flat file either at Application server layer or HSM systems (Hierarchical Storage Management) or other third-party provider's storage system like OpenText using SAP Archivelink. This storage system stores the processed archive file after a delete program is successfully completed.

Advantages:

- This is traditional data archiving and valid for all traditional databases including SAP HANA
- No separate license required

Disadvantages:

- Only applicable for standard objects otherwise it requires custom development
- Main use case is SAP ECC or S/4HANA
- Slower retrieval

## 4 SUMMARY

SAP HANA Data Tiering is complex in nature and no 'one size fits all'. It utilizes different mechanism to archive data to cold and warm storage based on the HANA suites (whether Native, ECC or BW). Also, it depends on the price and complexity. This document summarizes each options and give a holistic overview of the processes.

## 5 ACKNOWLEDGMENTS

I would like to thank anonymous reviewers for the comments and suggestions.